\documentclass{phostproc}

\usepackage{flushend} 
\usepackage{color}
\newcommand{\Msolar}{\mbox{\,$\rm M_{\odot}$}}        

  \newcommand{\Teff}{\mbox{\,\em T$_{\rm eff}$}}         

  \def\simge{\mathrel{\raise1.16pt\hbox{$>$}\kern-7.0pt
    \lower3.06pt\hbox{{$\scriptstyle \sim$}}}}           
  \def\simle{\mathrel{\raise1.16pt\hbox{$<$}\kern-7.0pt
    \lower3.06pt\hbox{{$\scriptstyle \sim$}}}}           

\title{Post-common envelope binary stars:
Radiative Levitation and Blue Large-
Amplitude Pulsators}
\author{Conor Byrne$^{1,2*}$ \&
        Simon Jeffery$^{1,2}$}

\affiliation{$^{1}$ School of Physics, Trinity College Dublin, Dublin 2, Ireland \\
			 $^{2}$ Armagh Observatory \& Planetarium, College Hill, Armagh BT61 9DG, UK \\
			 $*$ Contact: conor.byrne@armagh.ac.uk}

\shorttitle{Origins of blue large-amplitude pulsators}
\shortauthors{Conor Byrne \& Simon Jeffery}

\abs{Following the recent discovery of a new class of pulsating star, the blue large-amplitude pulsators (BLAPs), pulsation stability analysis was carried out on evolutionary models of post-common envelope (CE) ejection stars of 0.3 and 0.46 solar masses. These models subsequently evolve to become a low-mass helium white dwarf and a core helium-burning extreme horizontal-branch star respectively. We investigate the effects of atomic diffusion, particularly radiative levitation, on the pulsation behaviour of the models. We find that when the models have effective temperatures comparable to those of BLAPs, the inclusion of radiative levitation allows sufficient enhancement of heavy metals to produce opacity-driven fundamental mode pulsations with periods similar to those observed in BLAPs.}

\begin{document}

\maketitle

\section{Introduction}

The discovery of a new class of pulsating star was announced about 18 months ago \citep{Pietrukowicz17}. These stars, which have been dubbed `blue large-amplitude pulsators' (BLAPs), were found by the OGLE survey, a long-term variability study of the Galactic Bulge and the Magellanic Clouds \citep{UdalskiOGLE4}. BLAPs vary in brightness by 0.2--0.4 mag over time periods of between 20 and 40 minutes. Initially, the prototype object, OGLE-BLAP-001 was characterised as a $\delta$ Scuti type variable due to the shape of its light curve. Follow-up observations showed that the object had an effective temperature, \Teff, of about $33\,000\,$K and a surface gravity of $\log(g/\rm{cm}\,\rm{s}^{-1}) \approx4.6$. Combined with the large pulsation amplitude, this led to the conclusion that while the star can be compared to a $\delta$ Scuti variable and also to hot subdwarf variables, it is in fact an entirely different class of pulsating star. Further observations over the course of the OGLE survey found several other BLAPs, bringing the total of known BLAPs to 14. It was also found that the atmosphere of these stars is deficient in hydrogen, with OGLE-BLAP-001 having a surface helium mass fraction of about 0.52. Therefore, the presence of pulsations in such objects is perhaps unsurprising when one considers that opacity-driven pulsations are shown to be excited in many regions of the luminosity--\Teff\ plane when hydrogen is deficient, reducing its ability to damp the oscillations \citep{JefferySaio16}.

The measured values of $\log(g)$ and \Teff\ for OGLE-BLAP-001 place it below the Main Sequence, suggesting an interesting evolutionary origin, perhaps related to that of the hot subdwarf stars. Subdwarf B type stars are low mass, core helium-burning stars with $20\,000 \leq T_{\rm eff} \leq 40\,000\,$K and $5.4 \leq \log(g) \leq 6.0$. While most have hydrogen-rich surfaces, many also have helium rich surfaces. Like BLAPs, many hot subdwarfs show periodic brightness variations due to pulsations. p-mode hot subdwarf pulsators have periods of around 2--9 minutes and surface temperatures of $28\,000 \leq T_{\rm{eff}}\slash K \leq 35\,000$. These pulsations are driven by the $\kappa$-mechanism resulting from the presence of a large iron and nickel opacity bump. These elements are enhanced in the outer layers through the process of radiative levitation. These pulsations were predicted theoretically by \cite{Charpinet96,Fontaine03} to be caused by iron accumulation and were discovered observationally by \cite{Kilkenny97,Green03}. It was demonstrated by \citet{JefferySaio06b} that nickel opacity also played a crucial role in the driving of these pulsations.

Radiative levitation is a result of differential radiative forces on different ions in the stellar interior, with the force being related to their electron structure. High ionisation states of iron-group elements have a dense spectrum of atomic lines and absorb a large fraction of the incident radiation as result. This produces a net upward force on these elements at temperatures around $2\times10^5\,$K. 
Competing with this process is the downward diffusion of heavier elements, referred to as gravitational settling. Gravitational settling is due to an imbalance in the gravitational and electric forces acting on ions. The competition between these two processes typically leads to elements accumulating in regions where the forces arising from radiative levitation and gravitational settling are in equilibrium.

One of the widely accepted theories of hot subdwarf formation is the stripping of the hydrogen envelope of a red giant star by a low mass companion star in a common envelope ejection (CEE) \citep{Han02,Han03}. If this envelope ejection takes place near to the tip of the red giant branch, it is possible for the star to subsequently ignite helium in the core and become a low mass horizontal branch star. By modelling the pulsations in the envelope of a BLAP, \citet{Pietrukowicz17} concluded that one of the likely models for a BLAP would be a 0.3\Msolar\ star with a large degenerate helium core, providing a further similarity to hot subdwarfs.

Recent work has examined the role of atomic diffusion, especially radiative levitation, on the evolution of a post-common envelope star as it evolves from the red giant branch to the extreme horizontal branch (EHB) following the onset of helium burning in the core, where it becomes a hot subdwarf star \citep{Byrne18}. It was found that radiative levitation plays an important role in determining the surface composition of such stars, even before they reach the horizontal branch.

\begin{figure}
    \centering
    \includegraphics[width=0.4\textwidth]{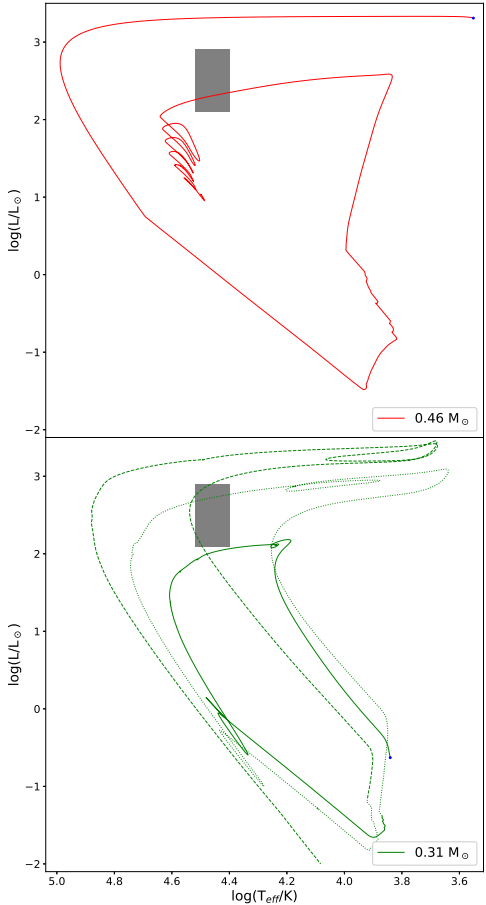}
    \caption{Hertzsprung-Russell diagram of post-CEE stars from red giant stars with a core helium mass of 0.46\Msolar\ (upper panel) and 0.31\Msolar\ (lower panel). For clarity, the `loops' in the diagram are indicated by a solid, dotted and dashed lines. The blue circles indicate the location of the star in the immediate aftermath of the envelope ejection. The grey shaded region indicates the typical luminosities and effective temperatures of BLAPs determined from observations.}
    \label{fig:evolution}
\end{figure}

Here we expand the analysis of these simulations to search for pulsations in our pre-subdwarf models and also investigate post-common envelope models of lower mass, to probe the extent of any instability strip which may exist, as well as the driving mechanism behind it. We then focus on specific models in these evolutionary sequences which have surface gravities and temperatures similar to those of BLAPs to test the possibility that BLAPs could be the direct progenitors of either hot subdwarfs or low mass white dwarfs.

\section{Method}

Evolutionary models of 0.46\Msolar\ and 0.31\Msolar\ post-CEE stars were produced using {\sc{mesa}} \citep{Paxton11,Paxton13,Paxton15,Paxton18}. The models started as a 1\Msolar\ zero-age Main Sequence stars. A standard solar metallicity was assumed ($\rm{Z}=0.02$, \citet{Grevesse98}), no mass loss (other than the common envelope ejection) was included, and convection was treated by considering only the Schwarzchild criterion. The stars were evolved until they reached the desired helium core mass on the Red Giant Branch, at which point a common envelope ejection is implemented.

As {\sc{mesa}} is unable to model such a dynamic phase of stellar evolution, the common envelope phase is replaced with a phase of very rapid mass loss ($\dot{\rm{M}}=10^{-3}\,\rm{M}_\odot\,\rm{yr}^{-1}$) until only a very low mass hydrogen envelope of $\sim 2\times10^{-3}\Msolar$ remains. These models were then allowed to evolve until becoming an extreme horizontal branch star (central helium mass fraction drops below 0.925) or a white dwarf ($L/\rm{L}_\odot=-2$). A more detailed description of the methods used and the input physics that is assumed is outlined in \citet{Byrne18} for the 0.46\Msolar\ stars and \citet{Byrne18b} for the 0.31\Msolar\ stars.

\section{Results}

Post-CEE evolutionary tracks for a model with a 0.46\Msolar\ and a 0.31\Msolar\ helium core are shown in Figure~\ref{fig:evolution}. These models evolve to become a core-helium burning hot subdwarf star and a low mass white dwarf respectively. From these evolutionary tracks, it was seen that these objects pass through a region of the luminosity-temperature domain in which BLAPs have been observed to reside by \citet{Pietrukowicz17}. Thus, it was decided that these should be investigated in further detail to probe their stability, to see if they could pulsate in a manner comparable to that of the BLAPs. Pulsation analysis of the {\sc{mesa}} models was carried out using the {\sc{gyre}} stellar oscillation code \citep{Townsend13}. Non-adiabatic mode analysis of the structure of the a stellar model can indicate if it is stable or unstable to pulsations when the sign of the complex component of the eigenfrequency is considered. Due to the high-amplitude nature of the pulsations, only the fundamental radial mode of pulsation was considered. Two sets of simulations are presented here. One set considers the evolution of the star without any diffusion processes, while the other includes the effects of atomic diffusion, particularly radiative levitation. Radiative accelerations on different ions in different cells within the star can be computed in {\sc{mesa}} following the modifications made by \citet{Hu11}.

\begin{figure*}
    \centering
    \includegraphics[width=0.64\textwidth]{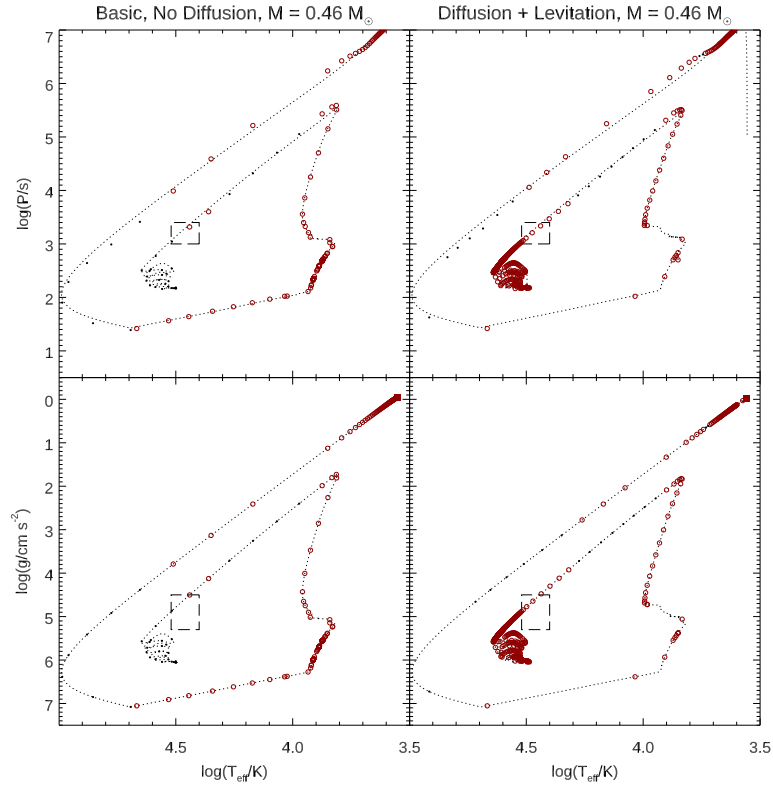}
    \caption{Pulsation stability during the course of evolution for a 0.46 \Msolar\ post-CE star. This figure is adapted from \protect\cite{Byrne18b}.}
    \label{fig:puls46}
\end{figure*}
\begin{figure*}
    \centering
    \includegraphics[width=0.64\textwidth]{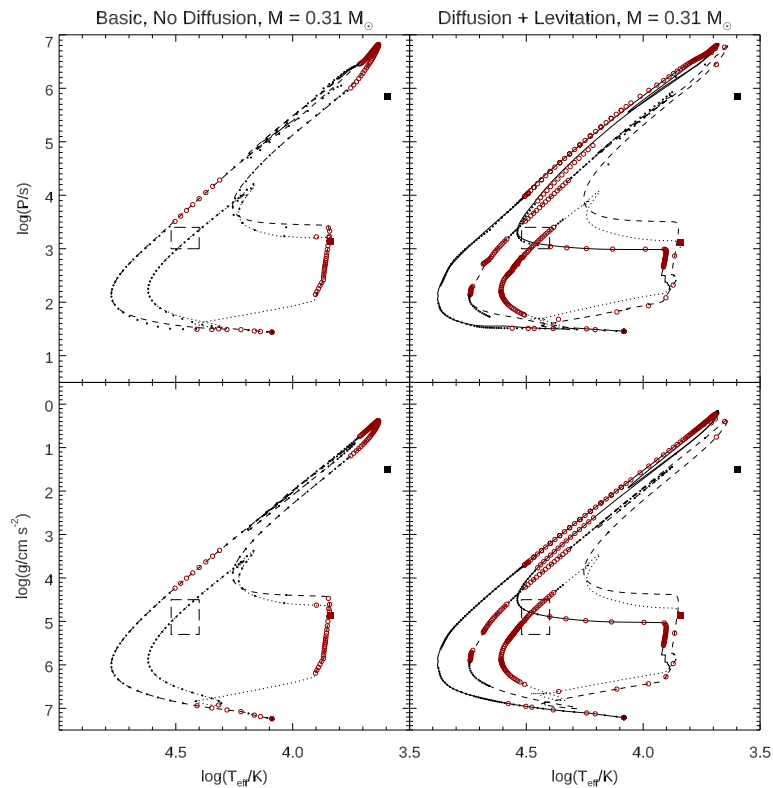}
    \caption{As in Figure~\ref{fig:puls46}, but for a 0.31\Msolar\ post-CE star. This figure is adapted from \protect\cite{Byrne18b}.}
    \label{fig:puls31}
\end{figure*}

Results of the analysis of pulsations the 0.46\Msolar\ post-CEE model is shown in Figure~\ref{fig:puls46}. Both surface gravity and fundamental mode period are plotted against effective temperature over the course of the evolution from the red giant branch to becoming a hot subdwarf. Individual dots represent stellar profiles from the {\sc{mesa}} evolution which were investigated via non-adiabatic analysis in {\sc{gyre}}. A small black dot indicates that the model was found to be stable in the fundamental radial mode, while the red open circles indicate models which are unstable. The dashed rectangles indicate typical gravities and periods found for BLAPs. One immediate observation is that the inclusion of atomic diffusion leads to significant regions of instability, most notably on the approach to the horizontal branch. This is not an unexpected result, given that pulsations in hot subdwarfs are known to be driven by z-bump opacity caused by accumulation of iron and nickel as a result of radiative levitation \citep[e.g.][]{Charpinet96}. The models produced here back up this finding, as it can be seen that the inclusion of radiative levitation vastly increases the amount of instability present in the vicinity of the horizontal branch. It is also worth noting that the evolution of the star in terms of luminosity/surface gravity and effective temperature is essentially unaffected by the inclusion or exclusion of diffusion processes.

It can also be seen that this instability is present when the star is still contracting to become a hot subdwarf, and that it can have a surface gravity comparable to that determined spectroscopically for BLAPs, with pulsation periods which also match. Comparing this to the rates of period change reported by \citet{Pietrukowicz17} shows that for this 0.46\Msolar\ model, the star is evolving at a rate which is an order of magnitude too fast to match the magnitude of the period changes. 

The results of the simulations on 0.31\Msolar\ models are shown in Figure~\ref{fig:puls31}. Once again, the impact of radiative levitation is clear, with much larger regions of instability when it is included. This lower mass star also passes through the portion of gravity-temperature space which is associated with BLAPs, and is seen to be unstable in the fundamental radial mode. this time, the rate of period change in this region is of the same order of magnitude as the observed rates for BLAPS $\left(\dot{\Pi} = \frac{\Delta P}{\Delta t} \sim 10^{-7}\,\rm{yr}^{-1}\right)$. 

The only issue with this evolutionary scenario is that both negative and positive rates of period change in the observations, while these models are contracting stars and only have a negative rate of period change for the fundamental mode. 

\begin{figure}
    \centering
    \includegraphics[width=0.4\textwidth]{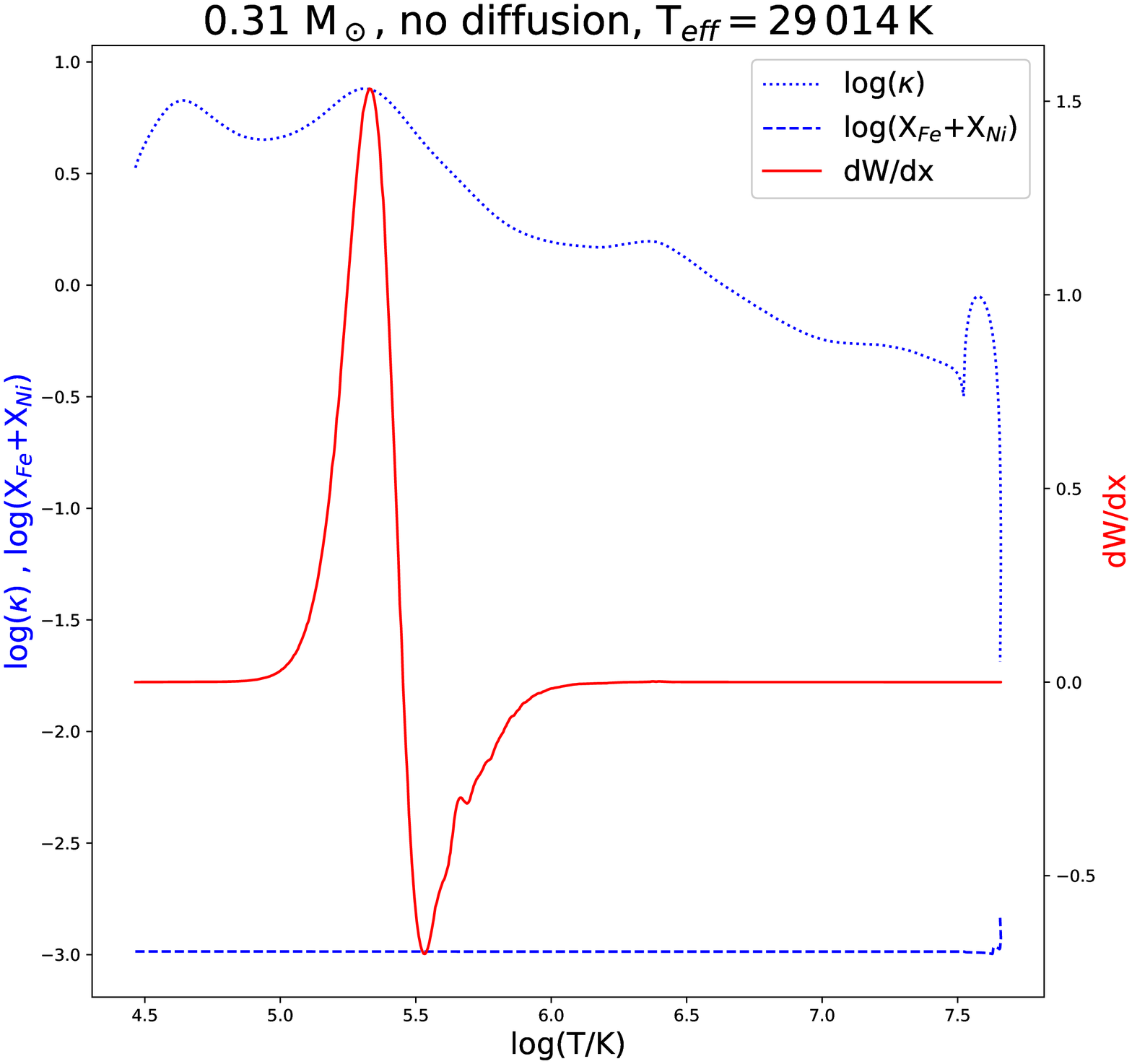}
    \includegraphics[width=0.4\textwidth]{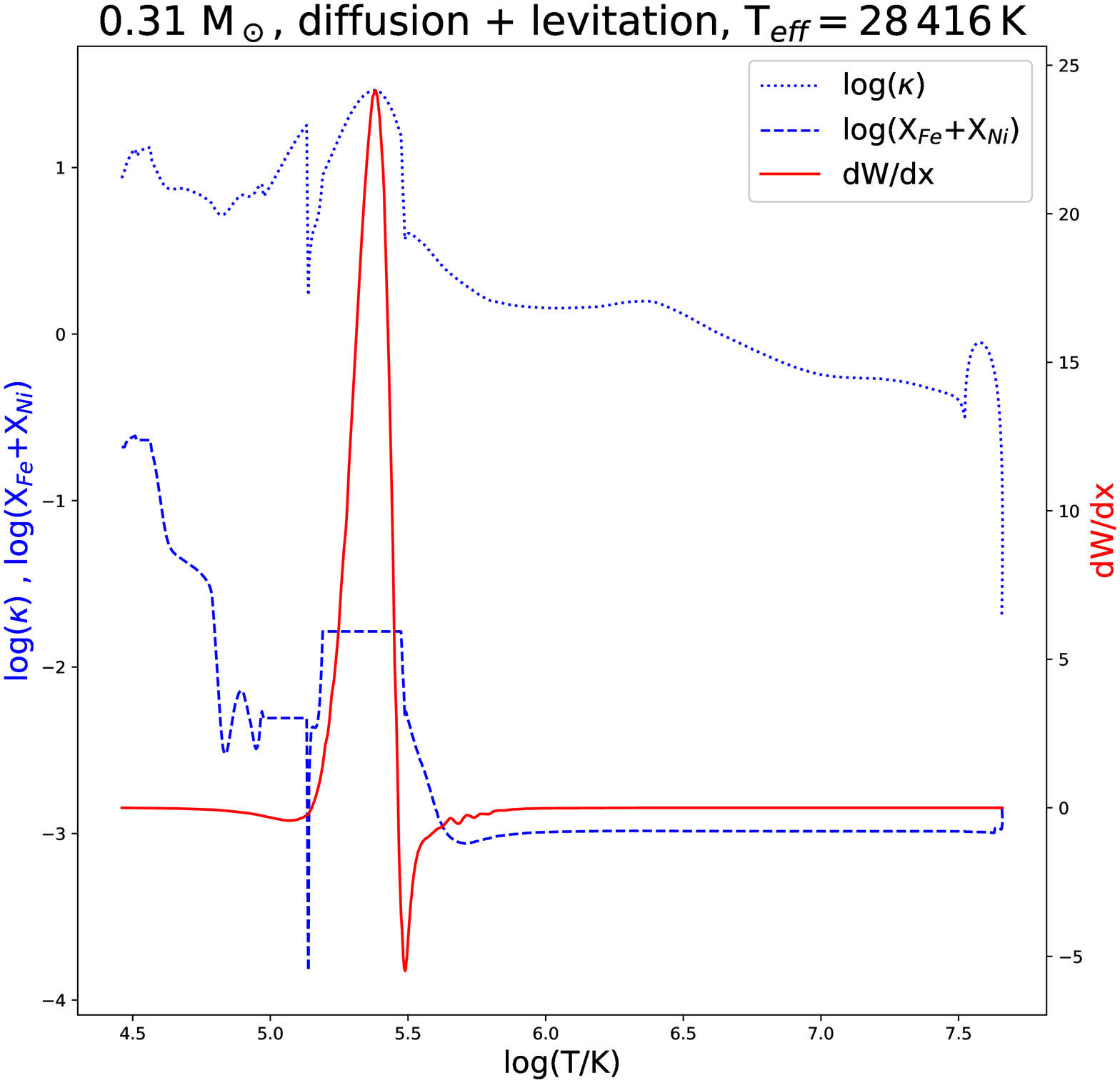}
    \caption{Opacity, heavy metal mass fraction and $d\rm{W}/dx$ of the radial fundamental mode as a function of temperature for selected models. Note the different scales used for the different variables, labelled on the left-hand and right-hand y-axes.  Upper panel: 0.31\Msolar\ model without diffusion. Lower panel: 0.31\Msolar\ model including the effects of atomic diffusion and radiative levitation.}
    \label{fig:drive}
\end{figure}

Selecting a model from the appropriate temperature regime allows probing of the driving mechanism of the pulsations. Figure~\ref{fig:drive} shows the combined iron and nickel mass fraction, opacity, and the derivative of the work function of the fundamental radial mode as a function of temperature. Note that $d\rm{W}/dx$ uses a different scale to the other two variables, indicated on the right hand axis. The model without diffusion (top panel) shows a correlation between the peak in the driving and the iron opacity bump at $2\times10^5\,$K. The surface value of $d\rm{W}/dx$ is negative however, indicating that this pulsation will not be driven. The lower panel shows a similar model, but with radiative levitation included. The enhancement of iron and nickel in the driving region is clearly visible, and the peak of $d\rm{W}/dx$ is approximately an order of magnitude larger in this instance, with the surface value being positive. This confirms that the driving mechanism for these pulsations is opacity driving as a result of radiative levitation of iron and nickel. 

\section{Conclusions}

Radiative levitation plays a crucial role in the driving of pulsations in low mass, post-CE stars. Accumulation of iron and nickel in the z-bump enhances the opacity in the driving region producing significant instability. In the case of both a 0.46\Msolar\ pre-subdwarf and a 0.31\Msolar\ pre-white dwarf, the accumulation of iron and nickel can lead to opacity driving pulsations, including in the fundamental radial mode. These pulsations are present when the star has an effective temperature comparable to that of BLAPs, with periods which are commensurate with observations. When considering the time spent by each star in the appropriate temperature range, and the corresponding rates of period change, the 0.31\Msolar\ model is the most likely candidate for a BLAP. This agrees with the proposals of \citet{Romero18}, who suggested that a low-mass pre-white dwarf could pulsate with periods that match BLAPs, if enhancement of iron and nickel as a result of radiative levitation was considered.  As these pre-white dwarfs are contracting stars, the rate of change of the fundamental mode is negative, which can only explain some of the BLAPs. The origin of BLAPs with a positive rate of period change remains unexplained. One possibility could be the expansion phase of the `loops' produced by off-centre helium ignition when a pre-subdwarf is approaching the horizontal branch (see Figure~\ref{fig:evolution}), if these occurred at slightly cooler temperatures than the specific model presented here. However, the timescale of these loops is likely to be too rapid to match the rate of period change observed in BLAPs.

This work presented here is described in more detail in \citet{Byrne18b}.  

\section*{Acknowledgments}
CB acknowledges funding from the Irish Research Council (Grant Number GOIPG/2015/1603) and thanks the participants at PHOST for various discussions throughout the conference. SJ acknowledges support from the UK Science and Technology Facilities Council (STFC) Grant No. 
ST/M000834/1.  The Armagh Observatory \& Planetarium is funded by the Northern Ireland Department for Communities.

\bibliographystyle{phostproc}
\bibliography{references.bib}

\end{document}